\documentclass[conference]{IEEEtran}
\IEEEoverridecommandlockouts

\usepackage[utf8]{inputenc}
\usepackage[T1]{fontenc}
\usepackage[hypertexnames=false,hidelinks]{hyperref}
\hypersetup{pdftitle={Multi-Agent AI Design of a Nanowatt Biomedical Low-Pass Filter from Topology to Layout}, pdfauthor={Danial Noori Zadeh, Devanshu Prasad, Mohamed B. Elamien}}
\usepackage{cite}
\usepackage{url}
\usepackage{booktabs}
\usepackage{array}
\usepackage{amsmath,amsfonts}
\usepackage{microtype}
\usepackage{graphicx}
\graphicspath{{media/}}
\usepackage[caption=false,font=footnotesize]{subfig}
\usepackage{stfloats}
\usepackage{enumitem}
\usepackage{orcidlink}
\usepackage{algorithm}
\usepackage{algorithmic}

\usepackage{xcolor}

\begin{document}

\title{Multi-Agent AI Design of a Nanowatt Biomedical Low-Pass Filter from Topology to Layout}

\author{\IEEEauthorblockN{Danial~Noori~Zadeh\,\orcidlink{0009-0001-8998-3343}, Devanshu Prasad\,\orcidlink{0009-0006-4503-6279}, and Mohamed B. Elamien\,\orcidlink{0000-0002-6528-0010}}
\IEEEauthorblockA{\textit{Department of Electrical and Computer Engineering, McMaster University}, Hamilton, ON, Canada\\
noorizad@mcmaster.ca, prasad2@mcmaster.ca, elamienm@mcmaster.ca}}

\maketitle

\begin{abstract}
We demonstrate a multi-agent framework that takes an analog circuit
from topology synthesis to layout. A researcher agent directs schematic
and layout agents, a four-tier memory stores the specification and lessons, and
separate agents review the designs. Deterministic tools provide
pass/fail simulation reports, symbolic analysis, graph-based netlist equivalence, parasitic
injection, and extraction with DRC and LVS in every optimization trial. A human
approves the specification and the layout plan. For a fourth-order, 250-Hz subthreshold biopotential low-pass filter, the researcher agent synthesized, through simulation-guided trial and error, a novel current-reuse architecture that shares a single dc bias branch between a super-source-follower biquad and a flipped-voltage-follower biquad.
The agents completed a DRC- and LVS-clean layout in 130-nm IHP SG13G2 CMOS that meets the specification in nominal post-layout simulation,
despite the noise--power trade-off and threshold-voltage sensitivity of
subthreshold operation. The filter's
input-referred noise over 0.5--200~Hz is 41.6\,\% below that of the
super-source-follower reference design at about equal core power.
\end{abstract}


\section{Introduction}
\label{sec:intro}

Automated flows take digital design from register-transfer level to manufacturable
layout~\cite{digital_openroad_ajayiOpenSourceDigital2019}. Analog
design has no comparable end-to-end flow because its performance depends on
layout parasitics and process variation, so a schematic that meets specification can still fail
after extraction or across process
corners~\cite{general_analogcad_gielenComputerAidedDesign2000,general_danial_genAI_review}.

\begin{figure}[!t]
\centering
\includegraphics[width=\columnwidth]{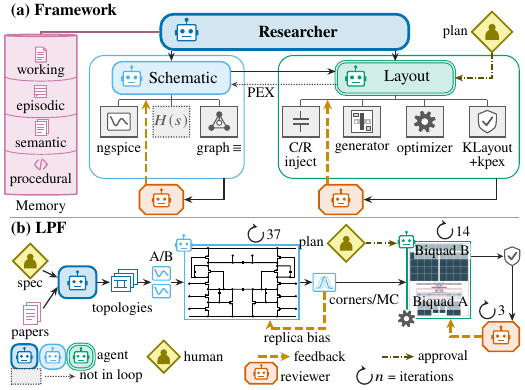}
\caption{(a) Framework; (b) LPF run. Dotted: not in this design's loop; $H(s)$: symbolic analysis; graph\,$\equiv$: netlist equivalence; C/R: parasitics and mismatch.}
\label{fig:arch}
\end{figure}

Large language model (LLM) agents assist at several stages of analog
design~\cite{EDA_agent_analogcoder_laiAnalogCoderAnalog2025,agentic_anaflow_ahmadzadehAnaFlowAgentic2025,agentic_analogcoderpro_laiAnalogCoderProUnifying2026,agentic_amsnetkg_shiAMSnetKGNetlist2025,agentic_ampagent_liuAmpAgentLLMbased2024,agentic_layoutcopilot_liuLayoutCopilotLLMPowered2025}.
Most of these design single blocks, such as amplifiers, at schematic
level, and LayoutCopilot needs a designer's instruction per step.
To our knowledge, none
targets subthreshold nanowatt design or feeds post-layout extraction
back into the layout of a complete filter.

A nanowatt biopotential low-pass filter (LPF) for wearable electrocardiogram
front ends~\cite{Prasad_Review} is a hard test case: with $g_m\propto I_D$ in weak inversion and a fixed cutoff, less current forces smaller capacitors and so more $kT/C$ noise (Section~\ref{sec:background}).
We ask whether LLM agents can carry a topology of their own from a written
specification to layout. The researcher agent synthesized the topology by trial and error: it combined techniques from published papers, simulated the candidates, and kept only the structural changes that the simulations supported (Section~\ref{sec:trail}). The
specification requires a cutoff $f_c$ of 245--255\,Hz, a maximum phase lag of at least 330$^\circ$ (the four-pole test), a gain of at most $-48$\,dB at 1\,kHz, a total harmonic distortion (THD) of at most $-40$\,dB at 175\,mV$_\text{pp}$ and 50\,Hz, an
input-referred noise (IRN) below 40\,$\mu$V$_\text{rms}$ (0.5--200\,Hz; reference design, two super-source-follower biquads: 49.98\,$\mu$V$_\text{rms}$), a filter-core power
below 50\,nW (reference design: 12.07\,nW), and a topology that
combines techniques from at least two of the ten published papers it lists. A human set the specification and reference design, sought small capacitors to save area, and approved the layout
plan; the agents ran all other design and verification steps.

Our contributions are:
\begin{enumerate}[leftmargin=*,nosep]
\item An end-to-end multi-agent framework for analog IC design, built on the SpiceXplorer platform~\cite{general_danial_newcas_score_shaping}, with a researcher agent that directs schematic-sizing and layout loops; an agent memory organized after the CoALA
  taxonomy~\cite{agentic_coala_sumersCognitiveArchitectures2024}; simulation,
  symbolic-analysis and layout tools built for agent use; and review by
  agents separate from the designer (Fig.~\ref{fig:arch}, Section~\ref{sec:method}).\footnote{The framework's tools are open source:
  \url{https://github.com/MacAnalog/spicexplorer-release}.}
\item A design run, autonomous between human approvals, from the
  specification to a DRC- and LVS-clean, extracted layout in IHP SG13G2 130-nm
  CMOS~\cite{tools_pdk_hermanVersatilityIHPBiCMOS2024}, with simulation in every design loop, including after extraction; the agents simulated 37 sizing points over 11 rounds and built 14 layout iterations, checked in 3 review rounds (Fig.~\ref{fig:arch}(b)).
  The agents A/B-tested the topology's two techniques and added a replica bias
  after process corners failed the cutoff specification
  (Section~\ref{sec:trail}). Against the reference design, the run lowered
  the IRN by 41.6\,\% at about equal power.
\item A fourth-order, 250-Hz subthreshold LPF that draws 11.91\,nW with an
  IRN of 29.19\,$\mu$V$_\text{rms}$ (0.5--200\,Hz) in post-layout
  simulation (Section~\ref{sec:circuit}).
\end{enumerate}

\section{Background}
\label{sec:background}
LLM agents assist at three stages of analog design~\cite{general_danial_genAI_review}.
For sizing, AnaFlow assigns specialized agents to interpret the topology and refine sizes in few simulations~\cite{agentic_anaflow_ahmadzadehAnaFlowAgentic2025}, and AmpAgent migrates published multi-stage amplifiers to new processes~\cite{agentic_ampagent_liuAmpAgentLLMbased2024}.
For topology, AnalogCoder writes and self-corrects circuit code from a library of sub-circuits~\cite{EDA_agent_analogcoder_laiAnalogCoderAnalog2025}; AnalogCoder-Pro couples topology generation with Bayesian-optimization sizing~\cite{agentic_analogcoderpro_laiAnalogCoderProUnifying2026}; and AMSnet-KG retrieves and assembles sub-circuits from a netlist knowledge graph~\cite{agentic_amsnetkg_shiAMSnetKGNetlist2025}.
AnalogSAGE adds stratified memory so that agents reuse simulation-grounded experience across op-amp tasks~\cite{agentic_AnalogSAGE}.
For layout, LayoutCopilot turns a designer's instructions into layout-tool commands~\cite{agentic_layoutcopilot_liuLayoutCopilotLLMPowered2025}.

The case study is hard because the noise, power and area specifications of a biopotential LPF conflict~\cite{Prasad_Review}.
In weak inversion each device's channel noise is shot noise, $S_{id}=2qI_D$, and $g_m=I_D/(nU_T)$, so the input-referred voltage-noise PSD, $2qI_D/g_m^2$, scales as $1/I_D$, while $f_c\propto g_m/C$ scales as $I_D/C$: at a fixed cutoff the integrated noise power scales as $\xi kT/C$, where $\xi$ is the circuit's excess noise factor, so lower noise costs capacitance and current.
The cutoff inherits the exponential dependence of $I_D$ on threshold voltage (Section~\ref{sec:trail}).
Stacking biquads in one branch reuses current but shares the 1.5-V headroom, which narrows the supply and temperature range.
Node impedances of tens of megaohms make the response sensitive to parasitics.
Prior subthreshold filters manage these trades with current reuse and follower-based biquads~\cite{Fifth_Column_Filter,Fourth_Column_Filter,Eight_Column_Filter}.

\section{Agentic Design Framework}
\label{sec:method}
\subsection{Agent Hierarchy}
\label{sec:hier}
The top-level agent, the researcher agent, holds the specification and
dispatches work to schematic and layout agents. Each sub-agent runs in its own context, starts
from its task and, for two schematic agents, a retrieved context (Section~\ref{sec:memory}) and
returns a short report; all agents run on the same LLM, Claude Opus~5.

Five agents design the schematic. A \emph{paper analyst} summarizes one paper
as a testable hypothesis: the mechanism on this filter, its predicted effect on
noise and power, and the A/B test that would refute it. A
\emph{variant runner} simulates candidate netlists in parallel, each batch with
the reference as a control. A \emph{sign-off verifier} signs off a candidate, a
\emph{schematic builder} draws it, and a fifth audits the
records for consistency. Each sub-agent has a fixed tool set; the fifth cannot write.

The layout agents start from the signed-off schematic netlist. A \emph{parasitic-budget agent}
converts the remaining pre-layout margin into per-net parasitic budgets. A
\emph{layout designer} writes a parameterized generator, which a \emph{layout reviewer} checks.

\subsection{Memory}
\label{sec:memory}
Memory follows the four CoALA types~\cite{agentic_coala_sumersCognitiveArchitectures2024},
and no store may exceed an agent's context window. \emph{Working memory} is a retrieved context of at most 20\,kB (specification, papers, lessons and recent runs),
re-assembled with the failure symptom as the query after each new
failure. \emph{Episodic memory} is a run log with one row per
simulation. \emph{Semantic memory}
holds lessons, each linked to its evidence, superseded rather than deleted when
wrong, and added to the next retrieved context.
For example, the agents recorded that the bias devices set the mismatch yield (Section~\ref{sec:trail}), and later cells enlarged those devices to raise it.
\emph{Procedural memory} is the measurement code, checks and agent definitions.

\subsection{Agent-Friendly Tools}
\label{sec:tools}
Each simulation is one call that returns a pass or fail result for each specification.
Simulation runs ngspice on fixed testbenches. The agents fixed a defect in their phase-lag measurement that reported a failing 320.7$^\circ$ as a passing 368.6$^\circ$; the corrected value disqualified the strongest candidate. Parasitic injection adds capacitance, resistance, a
threshold offset or leakage to the signed-off netlist and
re-simulates, giving each net a measured budget, so the layout designer moved biquad~A's vertical Metal2 route to the less capacitive Metal5.

A parameterized layout
generator, written as Python code in GDSFactory, gives the same GDS for the same parameters, which are the
optimizer's variables, and compares each layer with its mirror image, rejecting any build whose mismatch exceeds a tolerance. The check caught a 5-nm contact asymmetry. One call runs DRC, LVS and parasitic extraction (PEX) with KLayout and kpex,
and returns rule counts, unmatched nets and per-net capacitance. With each of the 33 generator parameters at both limits, DRC passed all 66 builds; adding LVS revealed shorted nets in two. A Nevergrad optimizer~\cite{tools_nevergrad_rapinOpenSourceEvolutionary2020} searched the 25 numeric generator parameters for minimum area, building, verifying and simulating a layout in every trial; its LVS caught a generator error that dropped capacitor-array plates. It also measured the cost of the human-approved bias dummy rows (5.6\,\% of the area), and the human reversed the approval.
A circuit graph turns a netlist into a bipartite graph of devices and nets, with devices labeled by type and model, and recognizes sub-blocks such as current mirrors~\cite{general_danial_analog_db}. A strict check, which also matches each device's model, showed that every sign-off netlist kept its connectivity after the transistor widths were rounded to values the PDK allows. A PMOS deliberately swapped to a high-voltage device passed the default check (type and polarity) but failed the strict one.

\emph{Symbolic analysis} builds on SLiCAP~\cite{tools_slicap_montagneStructuredElectronicsDesign2023} and SymXplorer~\cite{general_danial_symxplorer}: it derives exact transfer functions, simplifies them under numerically checked assumptions, and gives
pole-zero maps and per-device noise transimpedances. A human reviewer requested it after sign-off. The analysis
explained the poles and noise and matched the
pre-layout simulation within 0.03\,dB up to
1\,kHz. It also showed a pole pair turning real in 7 of 45 pre-layout PVT corners, which one-factor-at-a-time sweeps miss. The agents had sized the filter with a hand-derived biquad model, which misses this.

\subsection{Independent Review and Human Approval}
\label{sec:review}
In both loops, the designer and the verifier are separate agents. The sign-off verifier re-netlists
and re-simulates each cell submitted for sign-off rather than reading the
designer's results; the layout reviewer rebuilds the GDS from the generator,
re-runs DRC, LVS and extraction, and checks parasitics against the budgets. Neither may fix a
design, and the reviewer refuses to review a generator it wrote. Each finding
proposes a fix, usually to one generator parameter. The layout designer stops
for human approval of its plan; the human's four decisions included accepting a phase miss at one corner.

\subsection{Feedback Loops and Decisions}
\label{sec:trail}
The researcher agent evaluated the candidate topologies it had drawn from the ten papers in the specification~\cite{tian2023,fvf-2nd,ssf-33mhz,gmc-compact,Fifth_Column_Filter,Fourth_Column_Filter,selfcomp-gain,follower-63nw,biodevices,gmc-4p6nw}. In simulation, merging biquad~B's bias source and feedback device into one transistor cut core power by 66\,\% and drawn capacitance by 58\,\% against the unmerged stacked candidate, for only 1.31~$\mu$V$_\text{rms}$ more IRN, because it removed a bias-only noise source; the agents kept the merged device. A rail-tied bridge gate left the bridge in triode ($-2.17$\,dB dc gain), so the agents biased it from the bias reference.

\emph{A/B tests.} At equal capacitance, branch stacking~\cite{gmc-compact} with the merged device,
on super-source-follower biquads~\cite{ssf-33mhz}, cut core power by 67.5\,\% and the IRN by 18.1\,\% against the unstacked reference topology, because it reduces the bias-device count from eight to two and so lowers $\xi$; a floating
differential capacitor~\cite{fvf-2nd} cut drawn capacitance by 45.1\,\% at equal response. The human's area goal drove the floating-capacitor test and the dummy-row cost measurement (Section~\ref{sec:tools}).

\emph{Replica bias.} The stacked design passed 1 of 22 corners mixing process, supply and temperature extremes, so the agents replaced the bridge's reference bias with a shared
three-device replica bias ($M_{\mathrm{r1}}$--$M_{\mathrm{r3}}$), cutting the max/min $f_c$ ratio across
ss/ff/sf/fs from 3--38$\times$ to 1.01--1.02$\times$.

\emph{Mismatch.} On an unstacked candidate, enlarging all devices cost 21$^\circ$ of phase and failed the 330$^\circ$ phase-lag specification;
enlarging only the bias devices raised the candidate's Monte Carlo yield
while the phase lag stayed above 330$^\circ$, so the agents enlarged only those.

\section{Results}
\label{sec:results}

\begin{figure}[t]
\centering
\includegraphics[width=\columnwidth]{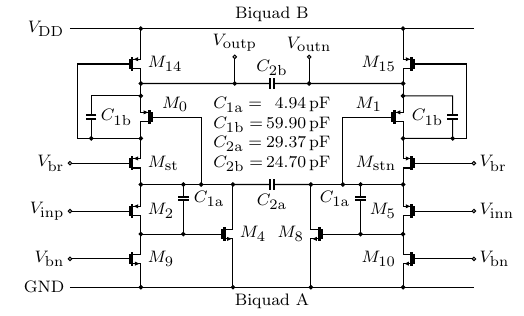}
\caption{Fourth-order 250-Hz filter core, one $V_{\mathrm{DD}}$-to-ground stack per half: biquads A and B,
bridge $M_{\mathrm{st}}/M_{\mathrm{stn}}$ biased at $V_{\mathrm{br}}$.}
\label{fig:schem}
\end{figure}

\begin{figure}[t]
\centering
\subfloat[$|H|$]{\includegraphics[width=0.485\columnwidth]{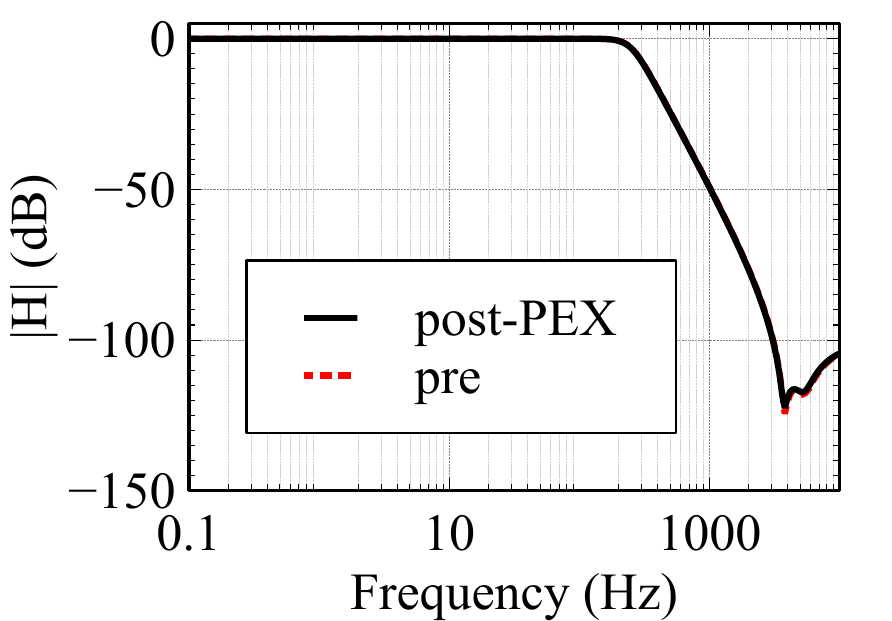}}\hfil
\subfloat[Phase, group delay]{\includegraphics[width=0.485\columnwidth]{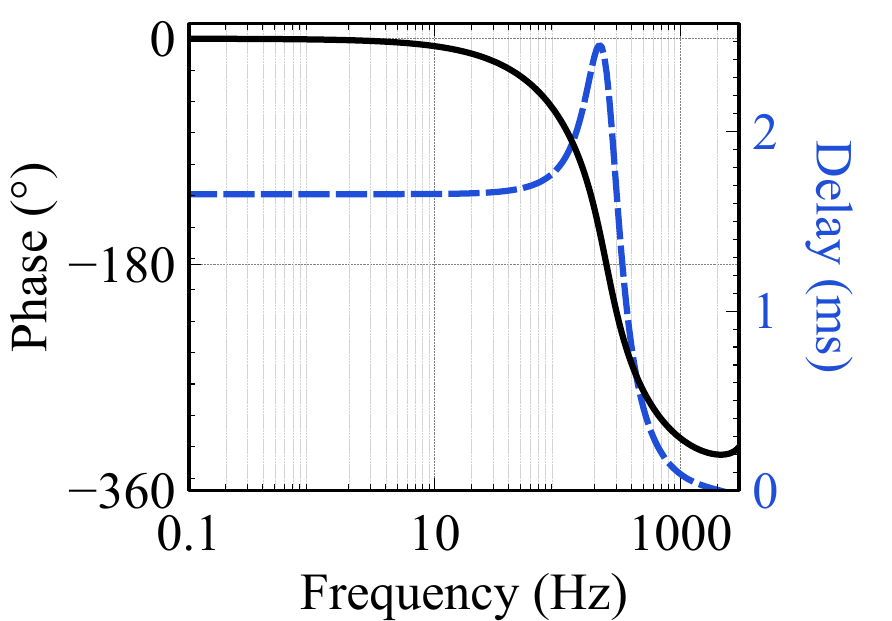}}\\[-2pt]
\subfloat[Input-referred noise]{\includegraphics[width=0.485\columnwidth]{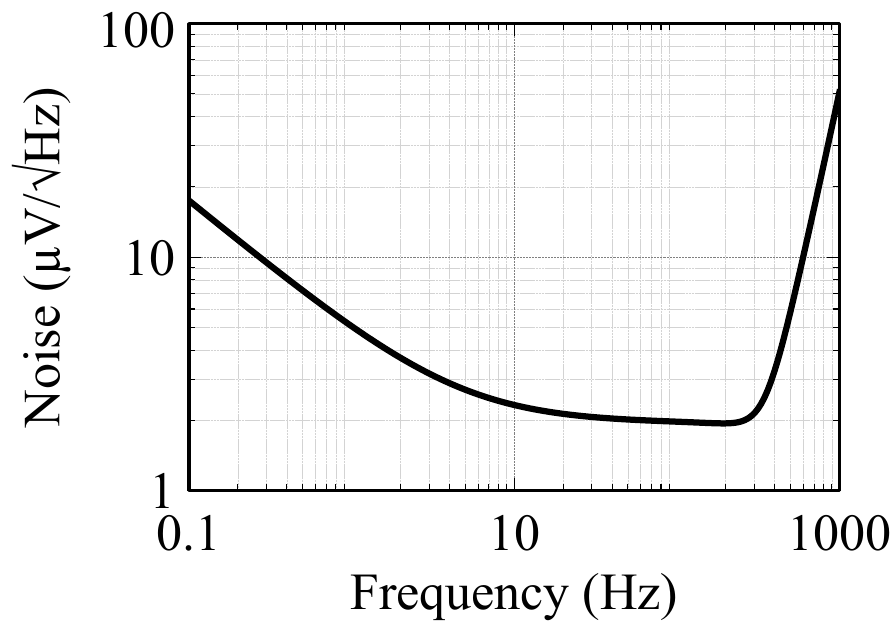}}\hfil
\subfloat[THD, 50\,mV$_\mathrm{pp}$]{\includegraphics[width=0.485\columnwidth]{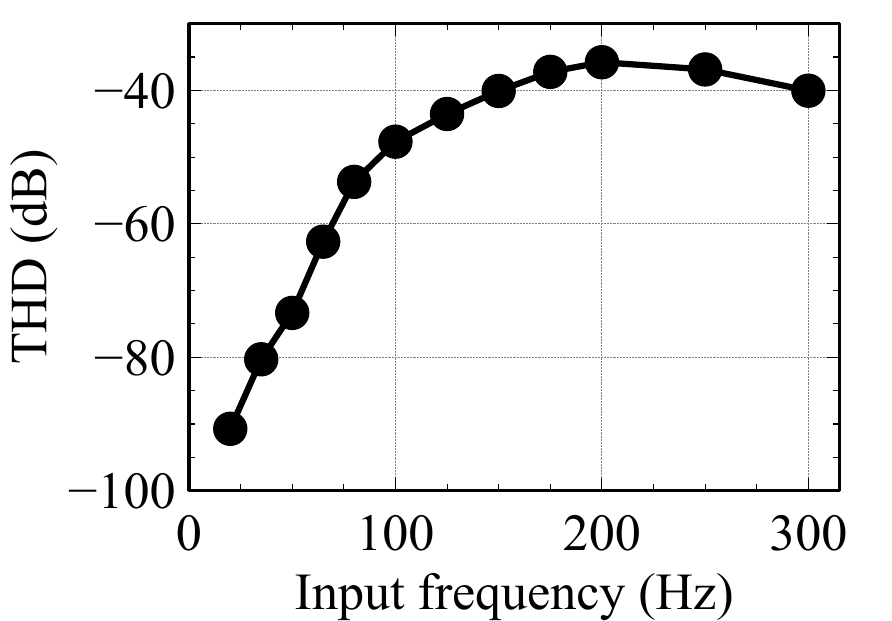}}
\caption{Post-PEX simulation, tt, 27\,$^\circ$C, 1.5\,V.}
\label{fig:results}
\end{figure}

\begin{figure}[t]
\centering
\includegraphics[width=\columnwidth]{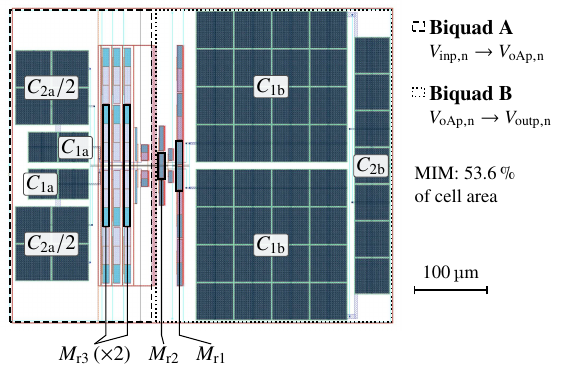}
\caption{Layout in IHP SG13G2: 0.228~mm$^2$, 53.6\,\% of it metal-insulator-metal (MIM) capacitors; DRC- and LVS-clean.}
\label{fig:layout}
\end{figure}

\begin{figure}[t]
\centering
\subfloat[CMRR, PSRR]{\includegraphics[width=0.485\columnwidth]{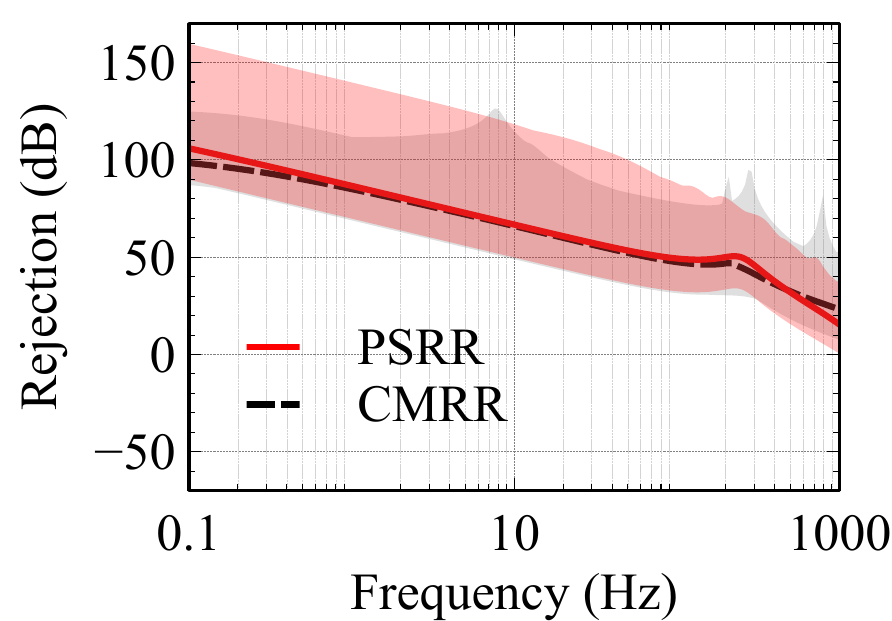}}\hfil
\subfloat[Monte Carlo $f_c$]{\includegraphics[width=0.485\columnwidth]{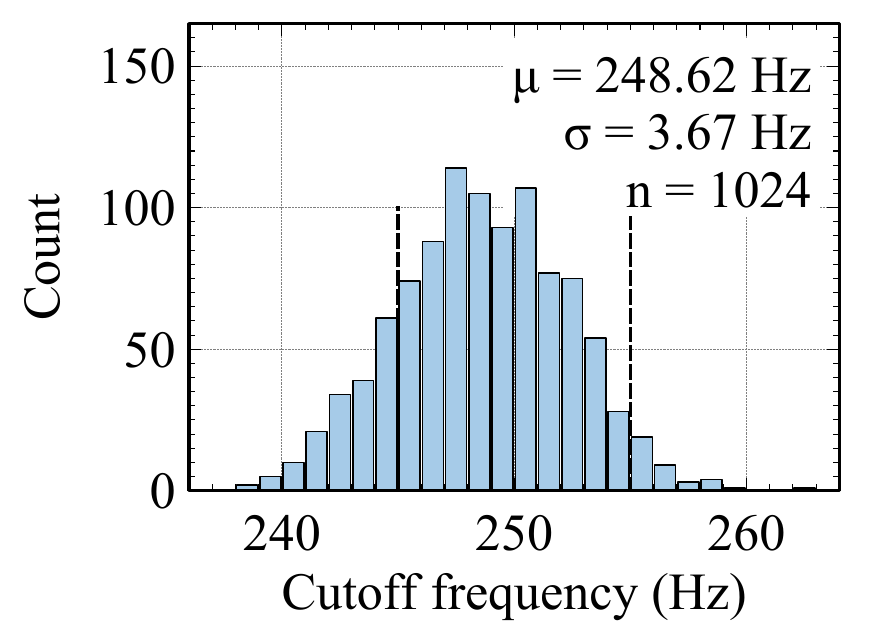}}
\caption{Post-PEX (a) CMRR and PSRR, mean and min--max, and (b) $f_c$ over 1024 Monte Carlo draws, spec limits dashed.}
\label{fig:robust}
\end{figure}

\begin{table*}[!t]
\centering
\caption{Comparison with Published Biopotential LPFs (This Work: Post-PEX Simulation, tt, 27\,$^\circ$C, 1.5\,V)}
\label{tab:comp}
\scriptsize
\setlength{\tabcolsep}{3.0pt}
\renewcommand{\arraystretch}{1.0}
\resizebox{\textwidth}{!}{%
\begin{tabular}{|l|c|c|c|c|c|c|c|c|c|c|c|}
\hline
 & \textbf{This Work} & \cite{Second_Column_Filter} & \cite{Third_Colum_Filter} & \cite{Fourth_Column_Filter} & \cite{Fifth_Column_Filter} & \cite{Sixth_Column_Filter} & \cite{Seventh_Column_Filter} & \cite{Eight_Column_Filter} & \cite{Ninth_Column_Filter} & \cite{Tenth_Column_Filter} & \cite{Eleventh_Column_Filter} \\
\hline
Sim./Meas. & Sim.\textsuperscript{a} & Meas. & Meas. & Meas. & Meas. & Meas. & Meas. & Meas. & Meas. & Meas. & Meas. \\
\hline
Process ($\mu$m) & 0.13 & 0.18 & 0.18 & 0.35 & 0.35 & 0.18 & 0.18 & 0.18 & 0.18 & 0.18 & 0.18 \\
\hline
Order & 4 & 4 & 5 & 4 & 4 & 5 & 5 & 4 & 2 & 3 & 2 \\
\hline
$V_{\mathrm{DD}}$ (V) & 1.5 & 0.5 & 1 & 0.9 & 1.5 & 1 & 1 & 0.5 & 1.2 & 1.2 & 0.8 \\
\hline
Power (nW) & 11.91 (13.90) & 13.5 & 453 & 4.26 & 5.25 & 350 & 41 & 3.69 & 38.4 & 16.3 & 5.6 \\
\hline
$f_c$ (Hz) & 248.66 & 240\textsuperscript{b} & 250 & 100 & 100 & 50 & 250 & 200 & 250 & 150 & 150 \\
\hline
$V_i$ (mV$_\mathrm{p}$) @ HD3$\,=-50$\,dB, $f$ (Hz)  & 87.5 @50 & 105 @80 & 50 @50\textsuperscript{c} & 29.5 @60 & 39 @60 & 43 @13\textsuperscript{d} & NR & 72 @50 & 270.5 @150 & 120 @40\textsuperscript{e} & 70 @90 \\
\hline
IIP3 (dBV$_\mathrm{rms}$) @ tones (Hz) & $-6.36$ @45/55 & $-7.9$ @190/200 & NR & $-14.0$ @50/60 & $-13.9$ @50/60 & NR & NR & $-8$\textsuperscript{f} @50/60 & 2.7 @125/150 & $-2.7$/$-7.3$ @30/45, 75/90 & $-3.9$ @30/45 \\
\hline
IRN ($\mu$V$_\mathrm{rms}$) & 32.9\textsuperscript{g} & 154.7\textsuperscript{h} & 340\textsuperscript{i} & 80.5\textsuperscript{g} & 39.38\textsuperscript{g} & NR & 134\textsuperscript{j} & 91.9\textsuperscript{g} & 82.1\textsuperscript{g} & 43\textsuperscript{j} & 35\textsuperscript{g} \\
\hline
DC gain (dB) & $-0.008$ & $-1.4$ & $-10.5$ & $-0.05$ & $-0.09$ & $-6$/$-8$ & $-7$ & $-5.6$ & $-0.18$ & 0.18 & $-0.1$ \\
\hline
DR (dB)\textsuperscript{k} & 65.5 & 53.6 & 40.3 & 48.27 & 56.9 & NR & NR & 54.9 & 67.35 & 65.9 & 63 \\
\hline
Area (mm$^2$) & 0.228 & 0.137 & 0.13 & 0.11 & 0.1 & 0.12 & 0.24 & 0.074 & 0.09 & 0.08 & 0.07 \\
\hline
FoM (fJ)\textsuperscript{l} & 6.4 (7.4) & 29.3 & 3485 & 41.1 & 18.7 & NR & NR & 8.3 & 33.0 & 18.4 & 13.2 \\
\hline
\end{tabular}}
\par\smallskip\parbox{\textwidth}{\scriptsize
\textsuperscript{a}Core power; in parentheses, with the ideal bias-reference branches. $V_i$ differential; IIP3 $-3.35$\,dBV peak per tone. \textsuperscript{b}Tunable 29--890\,Hz; DC gain $-3.2$ to $-1.1$\,dB over the range. \textsuperscript{c}HD3 $=-48.9$\,dB. \textsuperscript{d}THD $=-49.8$\,dBc. \textsuperscript{e}HD3 $=-51.9$\,dB. \textsuperscript{f}From 5.0\,dBm, 50\,$\Omega$ assumed.
\textsuperscript{g}Output noise over 1\,Hz--$2f_c$ divided by the passband gain. \textsuperscript{h}Input-referred over 1\,Hz--$2f_c$; referral method not stated. \textsuperscript{i}Noise band not reported. \textsuperscript{j}Integrated over 0.1--250\,Hz~\cite{Seventh_Column_Filter} and 0.5--300\,Hz~\cite{Tenth_Column_Filter}.
\textsuperscript{k}Dynamic range, recomputed: $\mathrm{DR}=20\log(V_{i,\mathrm{rms}}/\mathrm{IRN})$ at $\mathrm{HD3}=-50$\,dB ([\citen{Third_Colum_Filter}] and \cite{Eight_Column_Filter} report 50 and 48.5\,dB). \textsuperscript{l}$\mathrm{FoM}=P/(\mathrm{Order}\times f_c\times10^{\mathrm{DR}/20})$.}
\end{table*}

\subsection{Filter and Post-Layout Performance}
\label{sec:circuit}

The filter (Fig.~\ref{fig:schem}) is fully differential: biquad~A resembles a super-source follower~\cite{ssf-33mhz} and biquad~B a flipped voltage follower~\cite{fvf-2nd}, with all 15 devices in weak inversion; without isolated NMOS in SG13G2, both input followers are p-type to tie bulk to source. A p-type
common-gate bridge ($M_{\mathrm{st}}/M_{\mathrm{stn}}$) passes biquad~B's bias current into
biquad~A's output, so one 2.65-nA branch per half biases
both, extending branch stacking~\cite{gmc-compact} across biquads; $C_{2a}$ and $C_{2b}$ float between the halves~\cite{fvf-2nd}. A shared replica sets the bridge gate voltage,
so a mirror, not the threshold voltage, fixes the branch current. The
replica takes 33\,\% of the 7.942-nA core current but
holds $f_c$ across MOS process corners. Under mismatch, $f_c$ is
$248.62\pm3.67$~Hz (post-PEX, 1024 draws). Of 100 pre-layout draws of this filter (tt, 27\,$^\circ$C), 82\,\% meet every specification but THD (not run per draw).
Unlike a conventional fourth-order cascade, in which each biquad has its own bias current, this reuse couples the two biquads through a term $\kappa$ that brings the response close to Butterworth (Eq.~(\ref{eq:hs})).

Symbolic analysis (Section~\ref{sec:tools}) gives
$H(s)$, neglecting
$g_{ds}$ and device capacitances:
\begin{equation}
H(s)=\frac{g_{m,fa}\,g_{m,ia}\,g_{m,ib}\,(g_{m,br}+g_{m,fb})}
{D_A(s)\,D_B(s)+\kappa s^2},
\label{eq:hs}
\end{equation}
where $D_A$, $D_B$ are the isolated-biquad denominators and $i$, $f$, $br$
mark the input-follower, feedback and bridge devices of biquad $a$ or $b$.
The dc gain is unity for any sizing; finite $g_{ds}$ lowers it to
$-0.008$~dB (post-PEX). The coupling
$\kappa=2C_{1a}C_{2b}\,g_{m,br}\,g_{m,ib}$ moves the isolated $Q$ values
(2.10, 0.46) to 0.543 and 1.307 (full model, pre-layout; Butterworth: 0.541, 1.307; Fig.~\ref{fig:results}(a)).
Biquad~A's input followers ($M_2/M_5$) and bias sinks ($M_9/M_{10}$) carry 79\,\% of the pre-layout IRN power.

Capacitive extraction (Fig.~\ref{fig:layout})
moves $f_c$ from 249.77 to 248.66~Hz and leaves power and IRN
(29.19~$\mu$V$_\mathrm{rms}$, 0.5--200~Hz; Fig.~\ref{fig:results}(c)) within 0.02\,\%. At 50~Hz, post-PEX THD is
$-49.73$~dB at 175~mV$_\mathrm{pp}$, and the 50-Hz CMRR and PSRR over the 1024 draws (Fig.~\ref{fig:robust}(a)) average 52.2 and
53.4~dB (worst 35.6 and 36.6~dB).

\subsection{Comparison and Limitations}
\label{sec:comp}

The figure of merit (FoM) in Table~\ref{tab:comp} is the energy per pole per unit of dynamic range, so lower is better; the filter has the lowest, 6.4~fJ on core power and 7.4~fJ with the bias branches, against 8.3~fJ for~\cite{Eight_Column_Filter}. Its $V_i$ and IIP3, and so its DR and FoM, are
evaluated at $0.2f_c$, the low end of the $0.2$--$0.8f_c$ range in Table~\ref{tab:comp}.
Post-PEX THD at 50~mV$_\mathrm{pp}$ rises with input frequency
(Fig.~\ref{fig:results}(d)) because the filter slew-limits near $f_c$.
ADO-LLM~\cite{agentic_adollm_yinADOLLMAnalog2024} makes Bayesian-optimization sizing more efficient with LLM-proposed design points, but as a fixed-topology sizer it could not have found the A/B gains of Section~\ref{sec:trail}, because they required topology changes.
With MOS process, supply (1.40--1.65~V) or temperature (0--70\,$^\circ$C) varied one at a time, post-PEX $f_c$ stays in 246.00--251.52~Hz; 16 of the 45 full combinations meet every specification.

\section{Conclusion}\label{sec:conclusion}
We presented a multi-agent framework for analog IC design in which a
researcher agent directs schematic and layout agents. Through the design of a
nanowatt biopotential low-pass filter, we demonstrated the autonomous design of a
new circuit topology by AI agents, from specification to layout: the researcher
agent synthesized the topology by trial and error with simulator feedback.
Despite the noise--power trade-off and threshold-voltage sensitivity of
subthreshold design, the agents delivered a DRC- and LVS-clean layout that
meets the specification at nominal conditions. The agents worked autonomously
between human approvals of the specification and the layout plan, guided by deterministic tools, memory and review. Future work will place
symbolic circuit analysis in the agents' loop with the simulators, to test whether
agents can discover new design techniques rather than only combine and size known
ones, and will measure the filter in silicon.

\section*{Acknowledgments}
This work was supported by the Natural Sciences and Engineering
Research Council (NSERC) of Canada through its Discovery Grant (DG)
Program under Grant \mbox{RGPIN-2024-06826}.

\bibliographystyle{IEEEtran}
\bibliography{references}

\end{document}